Л. С. Синев, И. Д. Петров

# Температурный коэффициент линейного расширения (в интервале температур от 130 до 800 K) боросиликатных стекол, пригодных для соединения с кремнием в микроэлектронике

Представлены результаты обработки измерений ТКЛР и относительного удлинения боросиликатных стекол двух марок – ЛК5 и Borofloat 33. Термомеханическим анализатором TMA7100 было измерено удлинение образцов стекла в интервале температур от 130 до 800 K (от −143 до +526 °C). Относительная погрешность косвенных измерений ТКЛР и относительного теплового удлинения стекол обеих марок не превышает ±5 и ±3 % соответственно. Представлены полиномиальные уравнения, аппроксимирующие полученные результаты измерений в зависимости от температуры. Полученные результаты облегчат задачи моделирования характеристик приборов микросистемной техники, изготавливаемых с применением анодной посадки кремния на стекло, а также их можно использовать для оптимизации температурного режима соединения кремния со стеклом.
Ключевые слова: линейное расширение, ТКЛР, термомеханический анализ, боросиликатное стекло

**Leonid S. Sinev, Ivan D. Petrov**

# Linear thermal expansion coefficient (at temperatures from 130 to 800 K) of borosilicate glasses applicable for coupling with silicon in microelectronics

**Abstract**

Processing results of measurements of linear thermal expansion coefficients and linear thermal expansion of two brands of borosilicate glasses — LK5 and Borofloat 33 — are presented. The linear thermal expansion of glass samples have been determined in the temperature range 130 to 800 K (−143 to 526 °C) using thermomechanical analyzer TMA7100. Relative imprecision of indirectly measured linear thermal expansion coefficients and linear thermal expansion of both glass brands is less than ±5 % and ±3 % respectively. Polynomial equations, approximating temperature dependence of obtained measurements, are presented. The results will facilitate the modeling of the characteristics of the devices which are used in microsystems engineering and fabricated by anodic bonding of silicon to glass; they can also be used to optimize the temperature regime of silicon to glass bonding process.

В микромеханике широко применяется анодная посадка кремния на стекло. Для повышения точности моделирования характеристик приборов необходимы данные о температурной зависимости теплофизических параметров применяемых марок стекол в формульной записи полиномиального вида. Производители предоставляют такие данные по средним значениям в одном или нескольких интервалах температур. Однако этих данных недостаточно для точного моделирования характеристик приборов в широких интервалах рабочих температур. В представленной работе приведены полиномиальные аппроксимации результатов измерений температурных коэффициентов линейного расширения (ТКЛР) и относительного удлинения боросиликатных стекол двух марок — ЛК5 (Лыткаринский завод оптического стекла, Россия) и Borofloat 33 (Planoptik, Германия).

Для получения экспериментальных данных в интервале температур от 130 до 800 K (от −143 до +526 °C) использовали термомеханический анализатор TMA 7100 SII Nanotechnology. Прибор измеряет линейные размеры образца в условиях тепловых и механических нагрузок.

В работе измеряли по три образца длиной 20 мм для каждой марки стекла, результаты измерений затем усреднялись. Образцы устанавливали согласно рис. 1 в соответствии с инструкцией на термомеханический анализатор. Линейное приращение длины и длину образца при комнатной температуре измеряли с помощью зонда из кварцевого стекла, опирающегося на образец с силой 0,2 Н. Перед началом измерений термокамеру анализатора охлаждали. Последующий нагрев проводили со скоростью 10 К/мин. Значения температуры и линейного приращения длины регистрировались автоматически программным обеспечением термомеханического анализатора через равные промежутки времени. Для каждого образца было получено по 20 экспериментальных точек в интервале от 129 до 800 К с шагом не менее 35 К. По окончании измерений для каждого образца проводили программную коррекцию искажений, вносимых материалом колбы и зонда, с использованием эталона боросиликатного стекла SRM731.

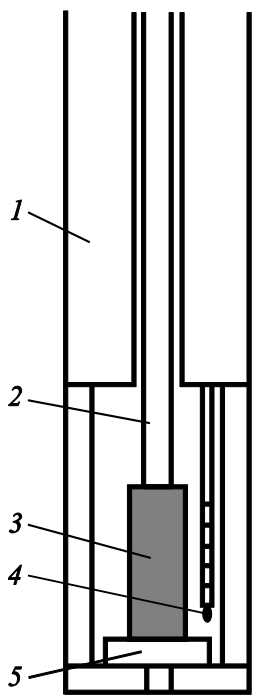

Рис. 1. Схема размещения измеряемого образца в термомеханическом анализаторе
*1* – колба; *2* – зонд; *3* – измеряемый образец; *4* – термопара; *5* – пьедестал

Значения ТКЛР $\alpha(T)$, К$^{-1}$, были рассчитаны следующим образом в соответствии с работой [1]:

$$\alpha(T) = \frac{\Delta L}{L_0 \Delta T}; \qquad (1)$$

$$T = \frac{T_i + T_{i+1}}{2};$$

$$\Delta T = T_{i+1} - T_i;$$

$$\Delta L = L(T_{i+1}) - L(T_i),$$

где $T$ – средняя температура между $i$ и $i + 1$ измеренными значениями, К;
$\Delta L$ – разница между $i + 1$ и $i$ измеренными значениями длины образца, м;
$L_0$ – исходная длина образца при комнатной температуре $T_0 = 293{,}15$ К, м;
$\Delta T$ – разница между $i + 1$ и $i$ измеренными значениями температуры, К;
$T_i$ – $i$-е измеренное значение температуры, К;
$L(T_i)$ – измеренное значение длины образца при температуре $T_i$, м.

Рассчитана зависимость относительного теплового удлинения образца от температуры $\varepsilon^T(T_i)$:

$$\varepsilon^T(T_i) = \frac{L(T_i) - L_0}{L_0}. \qquad (2)$$

Согласно [2] для данной измерительной установки пределы допускаемой относительной погрешности измерений линейных приращений длины составляют ±3 %, пределы допускаемой абсолютной погрешности измерений температуры равны ±1 °C.

Была проведена оценка относительных погрешностей косвенных измерений ТКЛР и относительного удлинения в точках их максимальных значений в рассматриваемом интервале температур.

Относительную погрешность ТКЛР согласно [3] рассчитывали по формуле

$$\frac{\delta\alpha}{\alpha} = \sqrt{\left(\frac{\delta(\Delta L)}{\Delta L}\right)^2 + \left(\frac{\delta(\Delta T)}{\Delta T}\right)^2 + \left(\frac{\delta L_0}{L_0}\right)^2}, \qquad (3)$$

где $\frac{\delta(\Delta L)}{\Delta L}$ – относительная погрешность измерений линейных приращений длины;

$\frac{\delta(\Delta T)}{\Delta T}$ – относительная погрешность измерений разницы температур;

$\frac{\delta L_0}{L_0}$ – относительная погрешность измерений длины образца при комнатной температуре.

Значения относительной погрешности измерений линейных приращений длины брали из данных, приведенных в [2].

Относительную погрешность измерений разницы температур определяли в соответствии с [3] по формуле

$$\frac{\delta(\Delta T)}{\Delta T} = \frac{\delta T \cdot \sqrt{2}}{\Delta T}, \qquad (4)$$

где $\delta T$ — абсолютная погрешность измерений температуры, взятая из [2], а в качестве величины $\Delta T$ принимали минимальное значение разницы температур из всей последовательности измерений в заявленном температурном интервале.

Относительную погрешность измерений длины образца при комнатной температуре с учетом данных по относительной погрешности измерений линейных приращений длины измерительной установкой [2] определяли в соответствии с [3] следующим образом:

$$\frac{\delta L_0}{L_0} = \frac{\delta(\Delta L)}{\Delta L} \frac{\Delta L}{\sqrt{2} L_0}, \qquad (5)$$

где в качестве величины $\Delta L$ принимали максимальное значение удлинений образца из всей последовательности измерений в заявленном температурном интервале.

В формуле (6) приведена оценка относительной погрешности измерения относительного удлинения образца согласно [3]:

$$\frac{\delta\varepsilon^T}{\varepsilon^T} = \sqrt{\left(\frac{\delta(\Delta L)}{\Delta L}\right)^2 + \left(\frac{\alpha\delta T}{\varepsilon^T}\right)^2 + \left(\frac{\delta L_0}{L_0}\right)^2}, \qquad (6)$$

где выражение $\frac{\alpha\delta T}{\varepsilon^T}$ введено аналогично описанному в [4] для учета зависимости $\varepsilon^T$ от температуры. В нем использованы расчетные значения $\alpha$ и $\varepsilon^T$, полученные из аппроксимации экспериментальных данных для температуры 800 K. Третий подкоренной элемент формулы (6) рассчитан по формуле (5), но в качестве $\Delta L$ принято

$$\Delta L = \varepsilon^T L_0. \qquad (7)$$

Относительная погрешность косвенных измерений ТКЛР стекол обеих марок, рассчитанная по формуле (3), не превышает ±5 %. Эта величина использована для отображения погрешности измерений на графиках рис. 2. Относительная погрешность косвенных измерений относительного удлинения стекол обеих марок, рассчитанная по формуле (6), не превышает ±3 %.

Функции, аппроксимирующие экспериментальные данные, получены методом наименьших квадратов. Данные функции имеют вид полиномов:

$$P(T) = a + b \cdot T + c \cdot T^2 + d \cdot T^3 + e \cdot T^4, \qquad (8)$$

где $a, b, c, d, e$ — коэффициенты полинома. Результаты аппроксимации экспериментальных данных ТКЛР и относительного температурного удлинения в диапазоне от 130 до 800 K полиномами не выше 4-го порядка сведены в таблицу. Полученные стандартные ошибки регрессий не превышают рассчитанных относительных погрешностей косвенных измерений, кроме аппроксимаций относительного удлинения в интервале температур 235 – 340 K. Кривые полученных зависимостей $\alpha(T)$ боросиликатных стекол Borofloat 33 и ЛК5 показаны на рис. 2.

Поскольку для обеих марок стекол есть только данные производителей по средним значениям на температурных интервалах, то сравним эти данные со средними значениями на тех же интервалах, рассчитанными по полученным в этой работе аппроксимациям.

Средний ТКЛР стекла ЛК5, по полученным в исследовании данным, в диапазонах от −60 до +20 °C и 20 – 120 °C совпадает с данными производителя [5].

А средний ТКЛР стекла Borofloat 33 в диапазоне 20 – 300 °C, по данным производителя [6], составляет $32{,}5 \cdot 10^{-7}$ °C$^{-1}$, по полученным в настоящей работе данным, это $31{,}2 \cdot 10^{-7}$ °C$^{-1}$.

**Таблица коэффициентов для уточненного расчета температурной зависимости теплофизических характеристик стекла на основе полиномиальной аппроксимации экспериментальных значений**

| Показатель | ЛК5 | | Borofloat 33 | |
|---|---|---|---|---|
| | $\alpha(T)$, K$^{-1}$ | $\varepsilon^T(T)$ | $\alpha(T)$, K$^{-1}$ | $\varepsilon^T(T)$ |
| $a$ | $1{,}123 \cdot 10^{-6}$ | $-7{,}494 \cdot 10^{-4}$ | $1{,}628 \cdot 10^{-6}$ | $-7{,}468 \cdot 10^{-4}$ |
| $b$ | $1{,}607 \cdot 10^{-8}$ | $9{,}57 \cdot 10^{-7}$ | $1{,}04 \cdot 10^{-8}$ | $1{,}451 \cdot 10^{-6}$ |
| $c$ | $-3{,}496 \cdot 10^{-11}$ | $8{,}46 \cdot 10^{-9}$ | $-2{,}198 \cdot 10^{-11}$ | $5{,}88 \cdot 10^{-9}$ |
| $d$ | $2{,}435 \cdot 10^{-14}$ | $-1{,}208 \cdot 10^{-11}$ | $1{,}398 \cdot 10^{-14}$ | $-8{,}426 \cdot 10^{-12}$ |
| $e$ | 0 | $6{,}246 \cdot 10^{-15}$ | 0 | $4{,}136 \cdot 10^{-15}$ |
| RSS | $25 \cdot 10^{-14}$ | $36 \cdot 10^{-11}$ | $5{,}9 \cdot 10^{-14}$ | $4{,}8 \cdot 10^{-11}$ |
| MSE | $17 \cdot 10^{-15}$ | $24 \cdot 10^{-12}$ | $3{,}9 \cdot 10^{-15}$ | $3{,}2 \cdot 10^{-12}$ |
| SER | $13 \cdot 10^{-8}$ | $4{,}9 \cdot 10^{-6}$ | $6{,}2 \cdot 10^{-8}$ | $1{,}8 \cdot 10^{-6}$ |
| *Примечания: a – e* – коэффициенты полинома (см. формулу (8)); RSS – сумма квадратов регрессионных остатков; MSE – среднеквадратичная ошибка; SER – стандартная ошибка регрессии. | | | | |

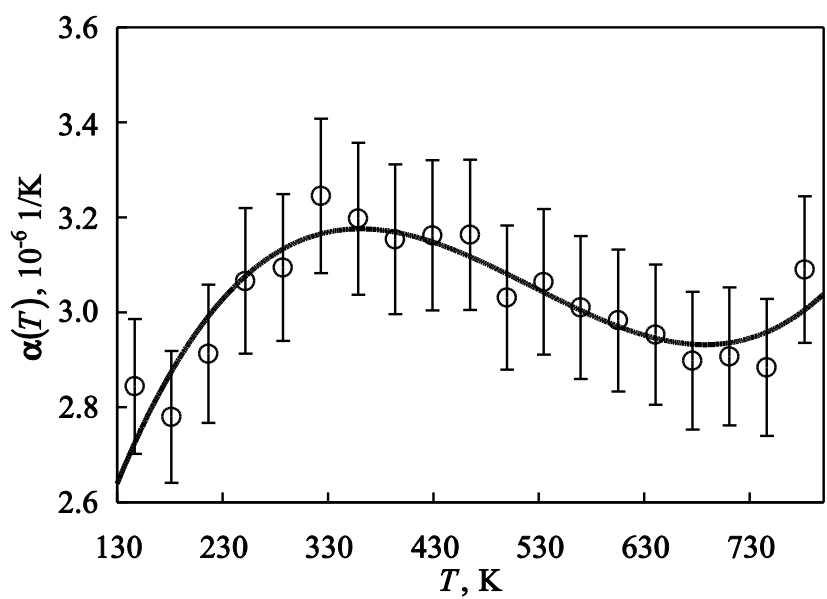

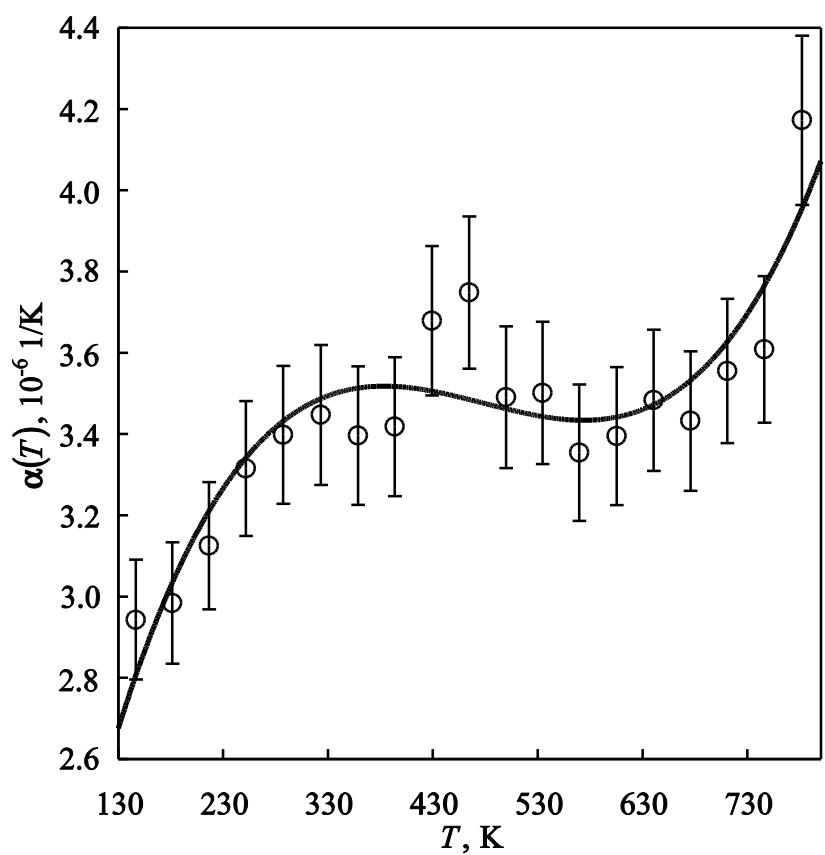

Рис. 2. Аппроксимация и экспериментальные значения температурного коэффициента линейного расширения стекол Borofloat 33 (*а*) и ЛК5 (*б*)

Таким образом, в данной работе были исследованы зависимости температурных коэффициентов линейного расширения и относительного теплового удлинения боросиликатных стекол ЛК5 и Borofloat 33 в интервале температур от 130 до 800 К (от −143 до +526 °C). Зависимости аппроксимированы полиномами не выше 4-го порядка. Коэффициенты полиномов для уточненного расчета температурной зависимости исследованных теплофизических характеристик стекла на основе полиномиальной аппроксимации экспериментальных значений приведены в таблице. Представление полученных данных в виде таблицы облегчит задачи моделирования характеристик приборов, изготавливаемых с применением анодной посадки кремния на стекло. Кроме того, результаты данной работы можно использовать для оптимизации температурного режима соединения кремния со стеклом и проектирования композиционных материалов [7] с использованием рассмотренных марок стекла.